\documentclass[reprint,prl,aps,twocolumn,showpacs,groupedaddress,amsmath,amssymb,floatfix]{revtex4-1}
\usepackage{graphicx}
\usepackage{dcolumn}
\usepackage{bm}

\begin{document}

\title{Unraveling Orbital Correlations via Magnetic Resonant Inelastic X-ray Scattering}

\author{Pasquale Marra}
\author{Krzysztof Wohlfeld}
\author{Jeroen van den Brink}
\affiliation{Institute for Theoretical Solid State Physics, IFW Dresden, Helmholtzstrasse 20, 01069 Dresden, Germany}
\date{\today}

\begin{abstract}
Although orbital degrees of freedom are a factor of fundamental importance in strongly correlated transition metal compounds, orbital correlations and dynamics remain very difficult to access, in particular by neutron scattering. 
Via a direct calculation of scattering amplitudes we show that instead \emph{magnetic} resonant inelastic x-ray scattering (RIXS) does reveal \emph{orbital} correlations. 
In contrast to neutron scattering, the intensity of the magnetic excitations in RIXS depends very sensitively on the symmetry of the orbitals that spins occupy and on photon polarizations.
We show in detail how this effect allows magnetic RIXS to distinguish between alternating orbital ordered and ferro-orbital (or orbital liquid) states.
\end{abstract}


\maketitle

\paragraph{Introduction}
Ever since the seminal work of Kugel and Khomskii~\cite{Kugel1982} in the 1980s it has been known that orbital degrees of freedom can play a crucial role in correlated transition metal compounds. 
Orbital ordering and orbital-orbital interactions are not only closely tied to magnetic ordering and magnetic interactions, but orbital degrees of freedom have also been proposed to be of direct relevance to spectacular phenomena such as colossal magnetoresistance in the manganites or superconductivity in the iron pnictides~\cite{Millis1995,Shimojima2011,Moreo2009}. 
Yet, the precise nature of correlated orbital states, being either of ordered or liquid type, and their existence in different materials is intensely debated, which to a large part is due to the fact that orbital correlations turn out to be very difficult to detect experimentally.
In fact, such experimental access would be of great help in unraveling the puzzling properties of many systems with orbital degrees of freedom, for instance the above mentioned iron-pnictide materials, where the type of the orbital ordering or its lack is heavily debated~\cite{Kruger2009,Bascones2010,Onari2010} or titanium and vanadium oxides where different theoretical scenarios --- a rather exotic orbital liquid phase~\cite{Khaliullin2000,Khaliullin2001}, or a classical alternating orbital-ordered state~\cite{Pavarini2004,Raychaudhury2007} --- have been proposed.

The experimental verification of orbital properties in correlated materials by neutron scattering is difficult because neutrons are almost not sensitive to the orbital symmetries of the ground state, in particular in orbital systems the angular momentum is quenched by the crystal field~\cite{Kim2011}.
Traditional x-ray diffraction, instead, is dominated by scattering from the atomic core electrons while the resonant x-ray diffraction~\cite{Elfimov1999,Benedetti2001}, particularly in the soft x-ray regime, the modern method of choice to detect orbital ordering, suffers from a very limited scattering phase space making Bragg scattering only possible for special orbital superstructures that have large spatial periodicities~\cite{Wilkins2003}.
There being few orbital-ordering related Bragg spots --- if at all --- leaves considerable room for controversies on the interpretation of experimental data~\cite{Benfatto1999,Wilkins2009,Beale2010}.

Recently, resonant inelastic x-ray scattering (RIXS) \cite{Groot1998,Ament2009,Haverkort2010,Ament2011} has been proven successful in measuring spin excitations in various cuprates~\cite{Braicovich2009a,Braicovich2009b,Bisogni2009,Schlappa2009,Guarise2010,LeTacon2011}, nickelates~\cite{Ghiringhelli2009a}, and even iron-based compounds~\cite{Hancock2010b}. 
Here we show in a general setting how the polarization dependent intensity of \emph{magnetic} RIXS directly provides an insight into the \emph{orbital} correlations in the ground state of correlated materials. 
In particular, we verify that RIXS discriminates between different orbital states, e.g., the alternating orbital (AO) order against the ferro-orbital (FO) order or the orbital liquid (OL) state. 
This method is applicable to any orbital-active material that has distinct dispersive spectral features in its spin structure factor $S({\bf k},\omega)$, for instance due to the presence of magnons arising from long-range magnetic ordering.
 
\paragraph{RIXS cross section}
RIXS is particularly apt to probe the properties of strongly correlated electrons, for instance in transition metal (TM) oxides~\cite{Ament2011}. 
With an incoming x-ray of energy $\omega_{\rm in}$ and momentum ${\bf k}_{\rm in}$ an electron is resonantly excited from a core level into the valence shell. 
At the TM $L_{2,3}$ edges this involves a $2p\rightarrow3d$ dipole allowed transition. 
In this intermediate state, the spin of the $2p$ core hole is not conserved, as the very large spin-orbit interactions strongly couple the spin and orbital momentum of the core hole. 
A spin flip in the core allows the subsequent recombination of the core hole with a $3d$ electron that has a spin opposite to the electron that was originally excited into the $3d$ shell. 
The energy $\omega_{\rm out}$ and momentum ${\bf k}_{\rm out}$ of the outgoing x-ray resulting from this recombination are then related to a spin excitation with energy $\omega=\omega_{\rm out}-\omega_{\rm in}$ and momentum ${\bf k}={\bf k}_{\rm out}-{\bf k}_{\rm in}$.

The magnetic RIXS cross section at a TM $L_{2,3}$ edge is in general~\cite{Ament2011, Haverkort2010}
\begin{equation}
\label{eq:crosssection}
 I_{\bf e}({\bf k},\omega)=\lim_{\delta\rightarrow0^+}{\rm Im }\langle0\vert\hat{O}^\dag_{{\bf k},{\bf e}}\frac1{\omega+E_0-H+\imath\delta}\hat{O}_{{\bf k},{\bf e}}\vert0\rangle,
\end{equation}
where $\mathbf{e}=\mathbf{e}^{\rm in}\cdot(\mathbf{e}^{\rm out})^{\dagger}$ is the tensor that describes the incoming and outgoing photon polarization, and $H$ is the Hamiltonian describing $3d$ valence electrons with ground state $\vert0\rangle $ and energy $E_0$. 
The Fourier transformed transition operator $\hat{O}_{{\bf k},{\bf e}}=1/\sqrt{N}\sum_{\bf j}\hat{O}_{{\bf j},{\bf e}}\exp(i{\bf k\cdot j})$ can be evaluated from the general expression for $\hat{O}_{{\bf j},{\bf e}}$ following the symmetry arguments in Ref.~\onlinecite{Haverkort2010} [cf.~Eqs.~(8)-(10)]
\begin{align}\label{eq:operator}
\hat{O}_{{\bf j},{\bf e}}=\sum_{d}\hat{n}_{{\bf j} d} \ 
\hat{\bf S}_{\bf j}\cdot {\bf W}_{\bf e}(d_{\bf j}),
\end{align}
where $\hat{\bf S}_{\bf j}$ are spin operators, $\hat{n}_{{\bf j} d}$ are number operators for electrons in the $3d$ orbitals on site ${\bf j}$, and where the vector amplitudes ${\bf W}_{\bf e}(d_{\bf j})$ depend on the orbital symmetry $d_{\bf j}$ of the ground state at site ${\bf j}$. 
Here each component of the vector ${\bf W}_{\bf e}(d_{\bf j})$ is \emph{a priori} different and thus each spin operator is multiplied by a distinct amplitude, which can be related to the fundamental x-ray absorption cross section%~\cite{Haverkort2010} %(cf.~Ref.~\onlinecite{Haverkort2010}) 
and therefore implicitly depends on the orbital occupancy $d_{\bf j}$ at site ${\bf j}$~\cite{Haverkort2010,Haverkort2010b,Groot1990}.

\begin{figure}[t!]
\includegraphics[width=1\columnwidth]{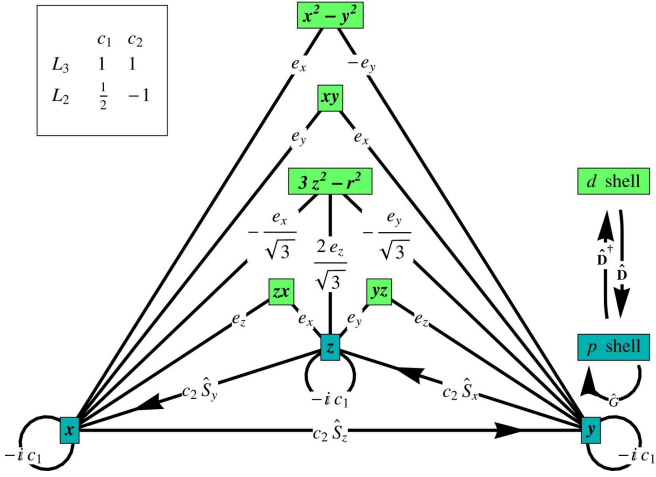}
 \caption{(color online)
Schematic representation of the RIXS operator $\hat{O}_{{\bf j}, {\bf e}}$ on a single site at the Cu$^{2+}$ $L_{2,3}$ edge.
To calculate the matrix elements of the operator between the same initial and final $3d$ orbital state, one needs to sum over all possible paths connecting them via a three step process, multiplying at every step as indicated in the figure: (i) the incoming polarization $e_\alpha$, (ii) $-\imath c_1$ or the spin operator $\pm c_2\hat{S}_\gamma$ [positive (negative) sign for steps along (opposite to) the direction of the arrows], and (iii) the complex conjugate of the outgoing polarization $e^*_\beta$ (constants $c_{1,2}$ depend on the edge). 
}
\label{fig:Diagram}
\end{figure}

\paragraph{Orbital dependence of RIXS operator}

As stated above the orbital dependence of RIXS amplitudes ${\bf W_e}(d_{\bf j})$ is generic to any orbital system. %~\cite{Haverkort2010,Haverkort2010b}. 
Nevertheless, to be explicit, we show how this dependence arises in the simple case of a Cu$^{2+}$ ion, i.e., with one hole in the Cu $3d$ orbital. 
The amplitudes ${\bf W_e}(d_{\bf j})$ can be evaluated using Eq.~(\ref{eq:operator}) as $W^\alpha_{\bf e}(d_{\bf j})\propto\langle d_{\bf j}\sigma_\alpha|\hat{O}_{{\bf j},{\bf e}}|d_{\bf j}\sigma_\alpha\rangle$ where $|d_{\bf j}\sigma_\alpha\rangle$ is the state with a hole in the $3d$ orbital with spin $\sigma$ along the $\alpha$ axis.
Since one needs here only to calculate the matrix elements of the operator $\hat{O}_{{\bf j}, {\bf e}}$ on single site states, this can be done just by applying the dipole and fast collision approximations to the Kramers-Heisenberg formula for RIXS~\cite{Veenendaal2006,Luo1993}, so that $\hat{O}_{{\bf j},{\bf e}}=\sum_{\alpha \beta}{e}_{\alpha \beta}\hat{D}^\dag_{{\beta},{\bf j}} \hat{G}_{\bf j}\hat{D}_{{\alpha},{\bf j}}$, where $\hat{D}_{\alpha {\bf j}}$ are the components of the dipole operator~\cite{Ament2011}, and $\hat{G}_{\bf j} \propto -\imath c_1+c_2\ \hat{\bf S}_{\bf j} \cdot\hat{\bf \Pi}_{\bf j}$ is the intermediate state propagator ($c_{1,2}$ are constants depending on the resonant edge, see Fig.~\ref{fig:Diagram}). 
The intermediate state transitions are expressed here by the operator $\hat{\Pi}_{\gamma \bf j}=\sum_{\alpha\beta}\epsilon_{\alpha\beta\gamma} p^{\dagger}_{ \alpha,{\bf j}} p^{}_{\beta,{\bf j}}$ where $\epsilon_{\alpha\beta\gamma}$ is the Levi-Civita symbol and $p^{\dagger}_{ \alpha,{\bf j}}$ the creation operator of the $2p$ core hole in the $p_{\alpha}$ orbital state.
This compact expression for the core hole propagator leads to the schematic representation of the operator $\hat{O}_{{\bf j},{\bf e}}$ on a single site in Fig.~\ref{fig:Diagram}. 

While the intermediate state propagator $\hat{G}_{\bf j}$ brings the spin dependence due to the spin-orbit coupling in the $2p$ core hole states, the dipole operators $\hat{D}_{\alpha{\bf j}}$ act in a different way depending on the orbital occupancy on site $\bf j$, so that the amplitude ${\bf W}_{\bf e}(d_{\bf j})$ strongly depends on the orbital symmetry of the ground state at each site. 
Since this dependence is merely due to the properties of the dipole transitions and to the spin-orbit coupling, it is indeed generic to any TM $L_{2,3}$ edge.

Having analyzed the inherent dependence of the scattering amplitudes ${\bf W_e}(d_{\bf j})$ on the \emph{single site} orbital occupancy, we now investigate how the operator $\hat{O}_{\bf j,e}$ in Eq.~(\ref{eq:operator}) acts on the orbital ground state of the \emph{bulk}. 
Hereafter, we consider  three different orbital ground states in a two-dimensional (2D) bipartite lattice (later we discuss a more general case): ferro-orbital (FO) order with the same $a$ orbital occupied on each site, alternating orbital (AO) order with $a$ ($b$) orbitals occupied on sublattice A (B), and orbital liquid (OL) ground state with the occupancies of $a$ and $b$ orbitals fluctuating similarly to the up and down spins in the spin liquid state.
Thus we obtain
\begin{equation} \label{eq:rixsamplitude}
\hat{O}_{{\bf j}, {\bf e}} =   
\Big[ \Big(\frac12 + \hat{T}^z_{\bf j} \Big) {\bf W}_{\bf e} (a) +
\Big(\frac12 - \hat{T}^z_{\bf j} \Big) {\bf W}_{\bf e} (b) \Big]
\cdot {\bf \hat{S}}_{\bf j},
\end{equation}
where the orbital pseudospin operator is $\hat{T}^z_{\bf j}=(\hat{n}_{{\bf j}a}-\hat{n}_{{\bf j}b})/2$.
Since $T^z_{\bf j}=1/2$ for all sites ${\bf j}$ in the FO state while $T^z_{\bf j}=\pm1/2$ for every other site in the AO state, the operator $\hat{O}_{{\bf j},{\bf e}}$ acts differently on different orbital ground states. 
Below we show how this feature affects spectra, by calculating the cross section using Eqs.~(\ref{eq:crosssection}) and~(\ref{eq:rixsamplitude}) for six ground states with different orbital and magnetic configurations. 

\paragraph{FM systems with AO order}
We consider a 2D ferromagnetic (FM) system with AO order (i.e., $\vert0\rangle=\vert\textrm{FM}\otimes\textrm{AO}\rangle$) with the spin interactions described by the \emph{effective} Heisenberg Hamiltonian $H=J\sum_{\langle{\bf i},{\bf j}\rangle}\hat{\bf S}_{\bf i}\cdot\hat{\bf S}_{\bf j}$ with negative exchange constant $J<0$. 
This spin-only Hamiltonian follows from a Kugel-Khomskii spin-orbital model when the interactions between orbital degrees of freedom generating the AO ground state are integrated out (see Part 1 of the Supplemental Material).

The spin wave (single magnon) excitation of such an ordered FM follows from the Holstein-Primakoff transformation for spins $\hat{S}^+_{\bf j}=\alpha_{\bf j}$, $\hat{S}^-_{\bf j}=\alpha^\dag_{\bf j}$ and $\hat{S}^z_{ \bf j}=1/2-\alpha^\dag_{\bf j}\alpha_{\bf j}$ with $\alpha^\dag_{\bf j}$ being bosonic creation operators: keeping the quadratic terms in $\alpha_{\bf j}$ and Fourier transforming one obtains the bosonic Hamiltonian $H=\sum_{\bf k}\omega_{\bf k}\alpha^\dag_{\bf k}\alpha_{\bf k}$ with spin wave dispersion $\omega_{\bf k}=2|J|(1-\gamma_{\bf k})$ where $\gamma_{\bf k}=(\cos k_x+\cos k_y)/2$.
Furthermore one has $\hat{T}^z_{\bf j}\vert0\rangle=\exp{(\imath{\bf Q}\cdot{\bf R}_{\bf j})}/2\vert0\rangle$ where ${\bf Q}=(\pi,\pi)$ is the AO ordering vector, so that following Eq.~(\ref{eq:rixsamplitude}) one obtains
\begin{align}\label{eq:operatorfmao}
 \hat{O}_{{\bf k},{\bf e}} &\vert  \textrm{FM} \otimes \textrm{AO} \rangle\! =\frac12 \Big\{ 
\big[ W^-_{\bf e} (a) + W^-_{\bf e} (b)  \big] \alpha^\dag_{\bf k} \nonumber \\
&\!+\!
\big[ W^-_{\bf e} (a) - W^-_{\bf e} (b)  \big] \alpha^\dag_{{\bf k} + {\bf Q}}
\Big\} \vert  \textrm{FM} \otimes \textrm{AO}  \rangle,
\end{align}
where $W^-_{\bf e}=W_{\bf e}^x-\imath W^y_{\bf e}$ are the amplitudes for the spin flip transition, which can be calculated for the simple case of a Cu$^{2+}$ ion (cf. Fig.~\ref{fig:Diagram}) or for any other TM ion (cf. Refs.~\onlinecite{Haverkort2010,Haverkort2010b}).
Using Eq.~(\ref{eq:operatorfmao}) and the spin Hamiltonian defined above, RIXS cross section can be directly calculated from Eq.~(\ref{eq:crosssection}) (cf. Fig.~\ref{fig:Bulk} and Part 2 of the Supplemental Material).
Due to the physical inequivalence of the two sublattices, the magnetic and orbital Brillouin zones are no longer the same, so that the backfolded branch of the magnon dispersion (pseudo optical branch in Fig.~\ref{fig:Bulk}) gains a finite intensity $\propto\vert W^-_{\bf e}(a)-W^-_{\bf e}(b)\vert^2$ [cf. Part 2 of the Supplemental Material and Eq.~(\ref{eq:operatorfmao})], as the spin flip amplitudes are different for orbitals $a$ and $b$.

\begin{figure}[t!]
\includegraphics[width=.95\columnwidth]{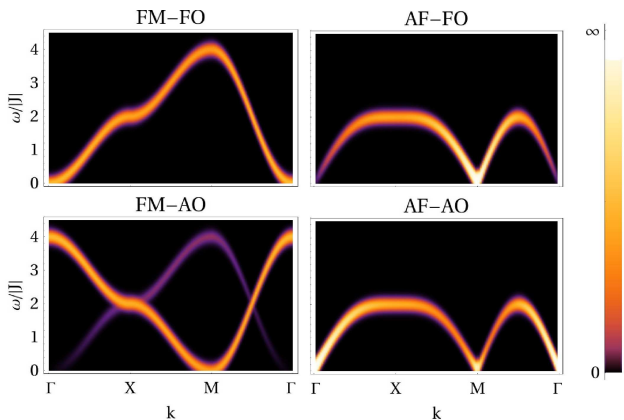}
\caption{(color online)
Magnetic RIXS cross section $I_{\bf e}({\bf k},\omega)$ for different magnetic (FM and AF) and orbital orders (FO and AO) along a high symmetry path in the Brillouin zone [where $\Gamma=(0,0)$, ${\rm X}=(\pi,0)$, and ${\rm M}=(\pi,\pi)$], averaged over incoming and outgoing polarizations.
The FO (AO) order is formed by the $x^2-y^2$ orbital ($x^2-z^2$ and $y^2-z^2$) while the spin quantization axis is in the $xy$ plane. 
The color scale is nonlinear, since intensities of the AF spectra diverge at $\rm M$, and at $\Gamma$ in the AF-AO case.
Spectra for the OL case (not shown) differ only quantitatively from the FO one.
}
\label{fig:Bulk}
\end{figure}

\paragraph{FM systems with FO order or OL state}
The above result stays in contrast with the 2D FM case with FO order ($\vert0\rangle=\vert\textrm{FM}\otimes\textrm{FO}\rangle$), for which one has $\hat{T}^z_{\bf j}\vert0\rangle=1/2\vert0\rangle$ for all sites ${\bf j}$.
Again using Eq.~(\ref{eq:rixsamplitude}) one obtains an equation for the operator $\hat{O}_{{\bf k},{\bf e}}$ and for the cross section in Eq.~(\ref{eq:crosssection}). 
In this case the orbital and magnetic Brillouin zones coincide since $W^-_{\bf e}(b)=W^-_{\bf e}(a)$ and there is no pseudo optical magnon branch in the RIXS cross section (see Part 2 of the Supplemental Material and Fig.~\ref{fig:Bulk}). 
Finally for the 2D FM case with an OL state ($\vert0\rangle=\vert\textrm{FM}\otimes\textrm{OL}\rangle$) with two fluctuating orbital states $a$ and $b$ the off-diagonal terms in $\hat{T}^z_{\bf j}\vert0\rangle$ lead to orbital excitations and therefore can be omitted from Eq.~(\ref{eq:rixsamplitude}), as we are interested only in pure spin excitations and not in coupled spin-orbital ones.
Again the orbital and magnetic Brillouin zones are identical and only the acoustic branch is detectable (see Part 2 of the Supplemental Material).

\paragraph{AF systems with AO order}
We consider a 2D antiferromagnet (AF) with AO order (i.e., $\vert0\rangle=\vert\textrm{AF}\otimes\textrm{AO}\rangle$) with the effective Heisenberg interaction between spins as in the FM case but with $J>0$.
Similarly to the previous case, the single magnon excitations are obtained by applying sequentially Holstein-Primakoff, Fourier, and Bogoliubov transformations and keeping only harmonic terms in bosonic operators $\alpha^\dag_{\bf k}$ and $\alpha_{\bf k}$, cf. Ref.~\cite{Kruger2009}, so that
\begin{align}\label{eq:operatorafao}
 \hat{O}_{{\bf k}, {\bf e}} & \vert  \textrm{AF} \otimes \textrm{AO} \rangle\! = 
\frac12 
\Big\{ 
\big[ W^+_{\bf e} (a)\! +\! W^-_{\bf e} (b) \big] u_{\bf k}\alpha^\dag_{\bf k}   \nonumber \\
&\!-\!\big[ W^+_{\bf e} (b)\! +\! W^-_{\bf e} (a) \big] v_{\bf k} \alpha^\dag_{\bf k}  \nonumber \\
&\!+\!
\big[ W^+_{\bf e} (a)\! -\! W^-_{\bf e} (b) \big] u_{{\bf k}+{\bf Q}} \alpha^\dag_{{\bf k}+{\bf Q}} \nonumber \\
&\!+\!\big[W^+_{\bf e} (b)\! -\! W^-_{\bf e} (a) \big] v_{{\bf k}+{\bf Q}}
\alpha^\dag_{{\bf k}+{\bf Q}} \Big\} \vert \textrm{AF} \otimes \textrm{AO}  \rangle,
\end{align}
with $W^+_{\bf e}=W_{\bf e}^x+\imath W^y_{\bf e}$ and where the Bogoliubov factors are defined as $u_{\bf k}=\sqrt{J/2\Omega_{\bf k}+1/2}$ and $v_{\bf k}={\rm sgn}(\gamma_{\bf k})\sqrt{J/2\Omega_{\bf k}-1/2}$, and the AF spin wave dispersion is $\Omega_{\bf k}=2J\sqrt{1-\gamma^2_{\bf k}}$.
This form of the operator in general leads to a nonvanishing intensity when ${\bf k}\rightarrow\Gamma$ as a result of the AO ordering, see Part 2 of the Supplemental Material and Fig.~\ref{fig:Bulk}. 
In the case of ideal AF $\Omega_{{\bf k}+{\bf Q}}=\Omega_{{\bf k}}$ so that in contrast to the $\vert\textrm{FM}\otimes\textrm{AO}\rangle$ case one can observe only one branch in the RIXS spectrum (although any corrections to the Heisenberg model for which $\Omega_{{\bf k}+{\bf Q}}\neq\Omega_{{\bf k}}$ will give rise to a pseudo optical branch in the spectrum, somewhat similar to the $\vert\textrm{FM}\otimes\textrm{AO}\rangle$ case).

\paragraph{AF systems with FO order or OL state}
Again the above result stays in contrast with the 2D AF case with FO order ($\vert0\rangle=\vert\textrm{AF}\otimes\textrm{FO}\rangle$) for which the RIXS operator has a simpler expression than Eq.~(\ref{eq:operatorafao}) since $W^\pm_{\bf e}(b)=W^\pm_{\bf e}(a)$. In a similar way intensities for the 2D AF case with OL state ($\vert0\rangle=\vert\textrm{AF}\otimes\textrm{OL}\rangle$) are obtained (see Part 2 of the Supplemental Material).
The intensity  vanishes in both cases when ${\bf k}\rightarrow\Gamma$ in agreement with Ref.~\onlinecite{Ament2009}, cf. Fig.~\ref{fig:Bulk} and Part 2 of the Supplemental Material.

\paragraph{Discriminating different orbital states}
As shown above for FM and AF systems, RIXS spectra can discriminate an AO against FO order or OL ground states (cf. Fig.~\ref{fig:Bulk}).
While in the FM case the pseudo optical magnon branch signals the onset of the AO order, in the AF case the intensity of magnons with momenta ${\bf k}\rightarrow\Gamma$ does not vanish in the AO case, contrarily to the FO and OL case.
This dependence is not due to distinct magnon dispersions for different orbital or electronic ground states~\cite{Wohlfeld2011,Tung2008,Ulrich2003}, but to the orbital dependency of magnetic RIXS amplitudes.

\begin{figure}[t!]
\includegraphics[width=.6\columnwidth]{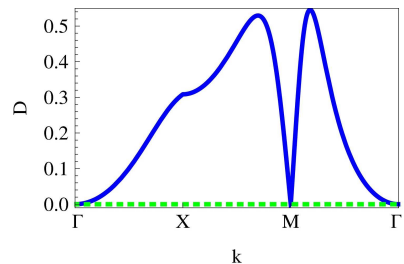}
\caption{(color online)
Circular dichroism $D=(I_{{\bf e}_L}-I_{{\bf e}_R})/(I_{{\bf e}_L}+I_{{\bf e}_R})$ for RIXS spectra intensities at $\omega=\Omega_{\bf k}$ as a function of transferred momentum ${\bf k}$ for the AF state, where ${\bf e}_L$ (${\bf e}_R$) is left (right) incoming circular polarization, for AO (FO and OL) state plotted with solid (dashed) line. 
}
\label{fig:Dichroism}
\end{figure}

Furthermore, circular dichroism of \emph{magnetic} RIXS intensities allows one to distinguish between different \emph{orbital} ground states, see Fig.~\ref{fig:Dichroism}. 
While for FM systems whether a circular dichroism is present depends on the symmetry of the orbital occupied, in the AF ones its presence only depends on the system translational symmetry.
Specifically, for $\vert\textrm{AF}\otimes\textrm{FO}\rangle$ (or $\vert\textrm{AF}\otimes\textrm{OL}\rangle$) systems, circular dichroism vanishes, while in the case of  $\vert\textrm{AF}\otimes\textrm{AO}\rangle$ order (for which the RIXS spin flip amplitude is finite for both orbitals forming the AO ground state, cf. Ref.~\cite{Ament2009}) the circular dichroism is nonzero (Fig.~\ref{fig:Dichroism}).

In fact, if there is an AO order in a magnetic system, translational symmetry is broken into two physically inequivalent sublattices. 
Consequently a pseudo optical branch in the magnon dispersion appears in the $\vert\textrm{FM}\otimes\textrm{AO} \rangle$ case. 
On the other hand, while a simple $\vert\textrm{AF}\otimes\textrm{FO}\rangle$ (or $\vert\textrm{AF}\otimes\textrm{OL}\rangle$) system is symmetric under the combination of time reversal and a discrete translation~\cite{Dresselhaus2008}, in the $\vert\textrm{AF}\otimes\textrm{AO}\rangle$ case the latter is broken. 
Macroscopically~\cite{Birss1966}, that means that the system is no longer symmetric under the combination of time reversal and translation. 
As a consequence, a finite circular dichroism appears, i.e., RIXS intensities (at fixed $\bf k$ and $\omega$) for left and right circular polarization of the incoming photon are no longer equivalent.

Although the actual values of the ${\bf W_e}(d_{\bf j})$ transition amplitudes depend on the orbital symmetry at each site,  differences in the RIXS spectra between the AO and the FO/OL ground states show up (cf. Fig.~\ref{fig:Bulk} and Fig.~\ref{fig:Dichroism}), as long as ${\bf W_e}(a)\neq {\bf W_e}(b)$. 
For this reason, the discrimination between different orbital states does not rely on the particular orbital occupancy on the \emph{single site}, but on the breaking of the translational symmetry caused by the onset of the AO orbital order. 

While other inelastic scattering methods have been theoretically proposed to detect orbital ordering~\cite{Ito1976,Veenendaal2008}, it should be stressed that, due to the onset of characteristic dispersion, the magnetic peaks in RIXS can, unlike, e.g., orbitons, be easily identified. 
Besides, as typically magnons interact weakly, quasiparticle peaks in RIXS spectra have sharp and well-defined line shapes which would rather not be obliterated by other low energy excitations (cf. Ref.~\cite{Ament2011}) and their dependence on the orbital ground state is thus very pronounced.

\paragraph{Conclusions}
We have shown in detail how ground state orbital correlations directly reflect themselves in magnetic resonant inelastic x-ray scattering (RIXS) intensities.
It follows that measuring the RIXS spectra at transition metal $L_{2,3}$ edges in correlated materials with orbital degrees of freedom and magnetic order, allows one to distinguish between different orbital ground states~\footnote{Although it seems not possible to distinguish between FO order and OL ground state in magnetic RIXS, typically in orbital systems the main question is whether the orbital ground state has AO order or is in the OL state~\cite{Khaliullin2000, Khaliullin2001, Pavarini2004, Raychaudhury2007} --- for which the proposed method is well suited.}.
This is possible because in magnetic RIXS the spin flip mechanism involves a strong spin-orbit coupling deep in the electronic core so that, unlike in inelastic neutron scattering, the magnetic scattering spectra strongly depend on the symmetry of the orbitals where the spins are in. 

The method proposed here is of direct relevance to 2D orbital systems, e.g., K$_2$CuF$_4$ or Cs$_2$AgF$_4$ with FM layers and predicted (but not yet explicitly verified) AO ordering~\cite{Hidaka1983, Wu2007, McLain2006}, as well as to three dimensional transition metal oxides with orbital degrees of freedom such as LaMnO$_3$, KCuF$_3$, LaTiO$_3$ or LaVO$_3$~\cite{Oles2005}. 
In particular, in Part 3 of the Supplemental Material we predict magnetic RIXS spectra for two different polytypes of KCuF$_3$ with distinct orbitally ordered ground states. 
\begin{acknowledgments}
We thank  V. Bisogni, M. Daghofer, M. W. Haverkort, C. Ortix and M. van Veenendaal for fruitful discussions.
K. W. acknowledges support from the Alexander von Humboldt Foundation. This research benefited from the RIXS collaboration supported by the Computational Materials Science Network (CMSN) program of the Division of Materials Science and Engineering, U.S. Department of Energy, Grant No. DE-FG02-08ER46540.
\end{acknowledgments}

%\bibliography{bib,supp}
%\bibliographystyle{apsrev4-1}

%Merlin.mbs v4.21 2009-07-09.
%

\newpage

\section{Supplemental Material}

\begin{table*}
%[h!b]
\begin{ruledtabular}
\begin{tabular}{ccc}
&FM     &AF    \\
\hline\noalign{\smallskip}
AO
& 
% FM AO:
$ \vert W^-_{\bf e}(a) + W^-_{\bf e}(b)\vert ^2
\delta (\omega-\omega_{\bf k})$ 
& 
% AF AO: 
$ \vert [ W^+_{\bf e}(a) + W^-_{\bf e}(b)] u_{\bf k}-[ W^+_{\bf e}(b) + W^-_{\bf e}(a)] v_{\bf k} \vert^2
\delta (\omega-\Omega_{{\bf k}})$ \\
& 
% FM AO (2):
$ +\vert  W^-_{\bf e}(a) -  W^-_{\bf e}(b)\vert ^2
\delta (\omega-\omega_{{\bf k}+{\bf Q}})$          
&
% AF AO (2):
$+\vert [ W^+_{\bf e}(a)- W^-_{\bf e}(b)] u_{{\bf k}+{\bf Q}}+[ W^+_{\bf e} (b)- W^-_{\bf e} (a)] v_{{\bf k}+{\bf Q}} \vert ^2 
\delta (\omega-\Omega_{{\bf k}+{\bf Q}})$
\\
\noalign{\smallskip}\hline\noalign{\smallskip}
FO      
&
%FO FM:
$\vert  W^-_{\bf e} (a)\vert ^2
\delta (\omega-\omega_{\bf k})$     
&
%FO AF:
$[\vert  W^+_{\bf e} (a) \vert ^2 + \vert  W^-_{\bf e} (a) \vert ^2 ] (u_{\bf k} - v_{\bf k})^2
\delta (\omega-\Omega_{\bf k})$
\\
\noalign{\smallskip}\hline\noalign{\smallskip}
OL
& 
% OL FM:
$ \vert  W^-_{\bf e} (a) +  W^-_{\bf e} (b)\vert ^2
\delta (\omega-\omega_{\bf k})$     
&
% OL AF:
$[\vert W^+_{\bf e}(a) + W^+_{\bf e}(b)\vert ^2 + \vert  W^-_{\bf e}(a) +  W^-_{\bf e}(b) \vert ^2 ](u_{\bf k} - v_{\bf k})^2 
\delta (\omega-\Omega_{\bf k})$
\end{tabular}
\end{ruledtabular}
\caption{
Magnetic RIXS cross sections $I_{\bf e} ({\bf k}, \omega)$ for three different orbital (AO, FO and OL states) and two different magnetic (FM and AF) ground states, see main text and supplemental materials for further details. Constant factors are omitted.}
\label{tab:1}
\end{table*}

The following supplementary information consists of three parts: in Part 1 we discuss the origin of the spin-only Hamiltonians which describes the spin-orbital problems in the main text of the paper, in Part 2 we give explicit equations for the magnetic RIXS cross sections as studied in the main text of the paper, and in Part 3 we calculate magnetic RIXS cross section of KCuF$_3$.

\subsection{1. Spin-only effective Hamiltonian}
\label{supp1}

In the main text we consider 2D systems with orbital order with the spin interactions described by the \emph{effective} Heisenberg Hamiltonian 
$$H =  J \sum_{\langle {\bf i}, {\bf j} \rangle} \hat{\bf S}_{\bf i}\cdot \hat{\bf S}_{\bf j}$$
with exchange constant $J$. 
Such a spin-only Hamiltonian follows from a Kugel-Khomskii spin-orbital model~\cite{Kugel1982} when the interactions between orbital degrees of freedom generating the AO ground state are integrated out (using e.g., mean-field decoupling).
In a typical AO and FO case with large crystal-field Jahn-Teller interactions this should always be possible~\cite{Wohlfeld2011}. 
In the OL case this might be questionable but the experimental results suggest that also in that case it is a valid approach~\cite{Oles2005}. 
However, if this is not the case, then in principle it is not clear what is the nature of the elementary excitations in such systems~\cite{Wohlfeld2011b}. 
In fact, so far there is only one case, closely related to the recently studied problem in Ref.~\cite{Wohlfeld2011b}, i.e., in the case of a FM-AO state with no Jahn-Teller interactions, in which the supposedly well-defined magnetic excitations dressed with the orbital ones should also give rise to two branches in the RIXS spectra. 
This is somewhat similar to the case described by Eq.~(\ref{eq:operatorfmao}) in the main text but with the pure magnetic excitations replaced by the ones dressed by orbitons (which would mean that each of the two branches visible in RIXS in Fig.~\ref{fig:Bulk} in the main text would also be visible but have also some large incoherent spectrum and potentially different periodicity). 

\subsection{2. Magnetic RIXS cross sections}
\label{supp2}

In Table \ref{tab:1} we give explicit equations for the magnetic RIXS cross sections for the three distinct ordered orbital ground states (AO, FO, and OL) and two distinct magnetically ordered ground states (FM and AF). 
This is calculated by substituting Eq.~(\ref{eq:operatorfmao}) [or Eq.~(\ref{eq:operatorafao})] into Eq.~(\ref{eq:crosssection}) in the main text for the FM (AF) case respectively. 
Note that results presented in Table \ref{tab:1} are then used in the main text \emph{inter alia} in Fig.~\ref{fig:Bulk}.

\subsection{3. Magnetic RIXS cross section of $\rm{KCuF}_3$}
\label{supp3}

In what follows we calculate RIXS cross section for the magnetic excitations in the so-called $A$-AF state (i.e., FM planes coupled AF along the $c$ direction) as stabilized in KCuF$_3$ below $T<T_N\sim38$K~\cite{Satija1980, Lake2000}. 
We consider here three different orbital states~\cite{Satija1980}: (i) $C$-AO state (i.e., AO planes coupled FO along the $c$ direction) as stabilized in the so-called (d)-type polytype of KCuF$_3$ below $T<T_S\sim800$K~\cite{Oles2005}, (ii) $G$-AO (i.e., isotropic 3D AO state) as stabilized in the so-called (a)-type polytype of KCuF$_3$ below $T<T_S\sim800$K~\cite{Oles2005}, and (iii) OL state (not realized in KCuF$_3$ but included for comparison). 
Furthermore we calculate the spectra for two different sets of $\{a, b\}$ orbitals forming the above discussed ground states: (i) the 3d$_{x^2-y^2}$ and 3d$_{3z^2-r^2}$ orbitals which can be preferred in the idealized case of vanishing interaction with the lattice (crystal field and Jahn-Teller interaction), and (ii) the 3d$_{x^2-z^2}$ and 3d$_{y^2-z^2}$ which is the set of occupied orbitals preferred by the interaction with the lattice and which is probably close to the one realized in KCuF$_3$~\cite{Oles2005}.

Similarly to the main text of the paper, we start with determining the spin wave excitations in the studied magnetic structure. These originate from the Heisenberg-like Hamiltonian
\begin{align}
 H= J_1 \sum_{\langle {\bf i}, {\bf j} \rangle || a,b} {\bf S}_{\bf i}\cdot {\bf S}_{\bf j}  + J_2 \sum_{\langle {\bf i}, {\bf j} \rangle || c} {\bf S}_{\bf i}\cdot {\bf S}_{\bf j},  
\end{align}
where the spin exchange constants $J_1 < 0$ and $J_2 > 0 $ lead to the onset of the $A$-AF ordered ground state.
The anisotropic structure of this spin-only Hamiltonian stems from the full Kugel-Khomskii spin-orbital Hamiltonian~\cite{Oles2005} when orbital degrees of freedom (generating one of the above mentioned orbital ground states) are integrated out. 
This means that the values of the spin exchange constants $J_1$ and $J_2$ depend on the orbital ground state. 

As in main text of the paper, the spin wave excitations (single magnon) can be calculated by applying sequentially the Holstein-Primakoff, the Fourier, and the Bogoliubov transformation and keeping only the harmonic terms in the bosonic operators $\alpha^\dag_{\bf k}$ and $\alpha_{\bf k}$. 
Thus we obtain the bosonic Hamiltonian $H = \sum_{\bf k} \varepsilon_{\bf k} \alpha_{\bf k}^\dag \alpha_{\bf k}$ with spin wave dispersion $\varepsilon_{\bf k} = \sqrt{A_{\bf k}^2 -B_{\bf k}^2}$, where $A_{\bf k} = 2|J_1| (1- \gamma_{\bf k}) + J_2$ and $B_{\bf k} = J_2 \mu_{\bf k}$ with $\gamma_{\bf k} = (\cos k_x + \cos k_y) /2 $ and $\mu_{\bf k} = \cos k_z$. 

Moreover, using Eq.~(\ref{eq:rixsamplitude}) in the main text we calculate the RIXS transition operator for %the $A$-AF (called $\textrm{S}(\bar{\bf Q})$ hereafter) and one of the orbitally ordered states mentioned above (called $\textrm{O}({\bf Q})$ below)
a ground state corresponding to a magnetic $\vert  \textrm{S}(\bar{\bf Q}) \rangle$ state and an orbital $\vert \textrm{O}({\bf Q})  \rangle$ one, with ordering vectors $\bf \bar{Q}$ and $\bf Q$ respectively
\begin{align}\label{eq:operatorsupp}
 & \hat{O}_{{\bf k}, {\bf e}}  \vert  \textrm{S}(\bar{\bf Q}) \otimes  \textrm{O}({\bf Q})  \rangle\! = \nonumber \\
\frac14 
\Big\{ 
&\!\big[W^+_{\bf e} (a)\! +\! W^-_{\bf e} (b) +  W^+_{\bf e} (b)\! +\! W^-_{\bf e} (a)  \big] 
\nonumber \\
&\times
(u_{\bf k} - v_{\bf k}) \alpha^\dag_{\bf k}   \nonumber \\
\!+&\! \big[ W^+_{\bf e} (a)\! + \! W^-_{\bf e} (a)- W^+_{\bf e} (b)\! - \! W^-_{\bf e} (b)  \big] 
\nonumber \\
&\times
(u_{{\bf k}+{\bf Q}} - v_{{\bf k}+{\bf Q}} ) \alpha^\dag_{{\bf k}+{\bf Q}}   \nonumber \\
\!+&\!
\big[W^-_{\bf e} (a)\! + \! W^-_{\bf e} (b) - W^+_{\bf e} (a)\! - \! W^+_{\bf e} (b)   \big] 
\nonumber \\
&\times
(u_{{\bf k}+{\bf \bar{Q}}} + v_{{\bf k}+{\bf \bar{Q}}} ) 
\alpha^\dag_{{\bf k}+{\bf \bar{Q}}}  \nonumber \\
\!+&\! \big[ W^+_{\bf e} (b)\! + \! W^-_{\bf e} (a)  - W^+_{\bf e} (a)\! - \! W^-_{\bf e} (b) \big] 
\nonumber \\
&\times
(u_{{\bf k}+{\bf Q} +{\bf \bar{Q}}} + v_{{\bf k}+{\bf Q} +{\bf \bar{Q}}} ) 
\alpha^\dag_{{\bf k}+{\bf Q}+{\bf \bar{Q}}} \Big\} \vert \textrm{S}(\bar{\bf Q}) \otimes  \textrm{O}({\bf Q})   \rangle,
\end{align}
where the Bogoliubov factors are defined as $u_{\bf k} = \sqrt{A_{\bf k}/(2 \varepsilon_{\bf k}) + 1/2}$ and $v_{\bf k} = {\rm sgn}(B_{\bf k})\sqrt{A_{\bf k}/(2 \varepsilon_{\bf k}) - 1/2}$.
Here the ordering vectors are defined through the equations $S^z_{\bf j} | \textrm{S}(\bar{\bf Q}) \rangle = \exp(\imath {\bf \bar{Q}}\cdot{\bf R_j})/2|\textrm{S}(\bar{\bf Q}) \rangle $ and $T^z_{\bf j} | \textrm{O}({\bf Q}) \rangle = \exp(\imath {\bf {Q}}\cdot {\bf R_j})/2   | \textrm{T}({\bf Q}) \rangle $ so that one has $ \bar{\bf Q} = (0, 0, \pi)$ for the $A$-AF phase and ${\bf Q} = (\pi, \pi, 0)$ [ ${\bf Q} = (\pi, \pi, \pi)$ ] for the $C$-AO ($G$-AO) phase and ${\bf Q} = (0,0, 0)$ for the OL phase. 
Note, however, that the above equations are valid also for other 3D spin and orbitally ordered phases with different ordering vectors and therefore can be used to study magnetic RIXS spectra for various other spin and orbitally ordered ground states.

\begin{figure*}[t!]
	\includegraphics[width=1.8\columnwidth]{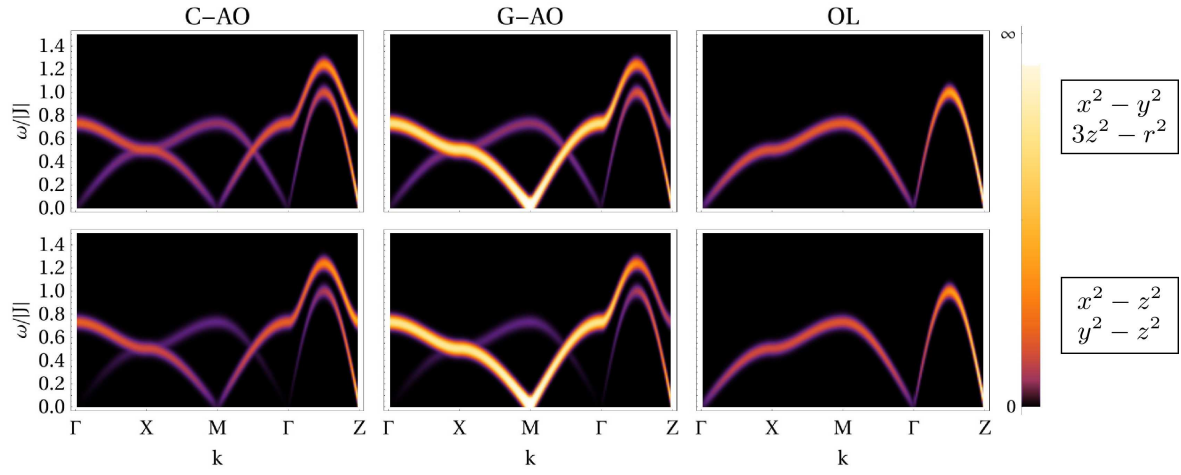}
     \caption{(color online)
Magnetic RIXS cross section $I_{\bf e} ({\bf k}, \omega)$ for KCuF$_3$ along a high symmetry path in Brillouin zone for three different orbital ground states (from left to right, $C$-AO, $G$-AO and OL state), formed by alternating $x^2-y^2$  and $3z^2-r^2$ orbitals (top row), and alternating $x^2-z^2$  and $y^2-z^2$ orbitals (bottom row), and with the spin quantization axis in the $xy$ plane. 
The cross section is averaged over incoming and outgoing polarizations and we assume that $|J_1/J_2| = 0.06$ in KCuF$_3$~\cite{Lake2000, Oles2005}. The high symmetry points in the Brillouin zone are defined as $\Gamma=(0,0,0)$, ${\rm X}=(\pi,0,0)$, ${\rm M}=(\pi,\pi,0)$, and ${\rm Z}=(0,0,\pi)$. The color scale is nonlinear, since intensities diverge at $\rm Z$ in every case, and at $\rm M$ in the $G$-AO case.
}
\label{fig:1}
\end{figure*}

Finally, using Eq.~(\ref{eq:operatorsupp}) above and Eq.~(\ref{eq:crosssection}) in the main text, we can calculate the RIXS cross section, that reads
\begin{align}\label{eq:intensity}
 & I_{{\bf e}} ({\bf k}, \omega) \propto \nonumber \\
&\!\big|W^+_{\bf e} (a)\! +\! W^-_{\bf e} (b) +  W^+_{\bf e} (b)\! +\! W^-_{\bf e} (a)  \big|^2 
\nonumber \\
&\times
(u_{\bf k} - v_{\bf k})^2 
\delta (\omega - \varepsilon_{\bf k})   \nonumber \\
\!+&\! \big| W^+_{\bf e} (a)\! + \! W^-_{\bf e} (a)- W^+_{\bf e} (b)\! - \! W^-_{\bf e} (b)  \big|^2 
\nonumber \\
&\times
(u_{{\bf k}+{\bf Q}} - v_{{\bf k}+{\bf Q}} )^2 
\delta (\omega -\varepsilon_{{\bf k}+{\bf Q}} )  \nonumber \\
\!+&\!
\big| W^-_{\bf e} (a)\! + \! W^-_{\bf e} (b) - W^+_{\bf e} (a)\! - \! W^+_{\bf e} (b)   \big|^2 
\nonumber \\
&\times
(u_{{\bf k}+{\bf \bar{Q}}} + v_{{\bf k}+{\bf \bar{Q}}} )^2 
\delta (\omega -\varepsilon_{{\bf k}+{\bf \bar{Q}}})  \nonumber \\
\!+&\! \big| W^+_{\bf e} (b)\! + \! W^-_{\bf e} (a)  - W^+_{\bf e} (a)\! - \! W^-_{\bf e} (b) \big|^2 
\nonumber \\
&\times
(u_{{\bf k}+{\bf Q} +{\bf \bar{Q}}} + v_{{\bf k}+{\bf Q} +{\bf \bar{Q}}} )^2 
\delta (\omega -\varepsilon_{{\bf k}+{\bf Q}+{\bf \bar{Q}}}),
\end{align}
and is shown in Fig.~\ref{fig:1} for the above mentioned three different orbital ground states and for two choices of the orbital sets forming these ground states.

While no clear signatures in the RIXS spectra allow one to distinguish between the two sets of \emph{single site} orbital occupancies $\{ x^2-y^2, 3z^2-r^2 \}$ and $\{x^2-z^2, y^2-z^2 \}$ (a subtle intensity shift between the optical and the acoustic branch is hardly visible), differences between the $C$-AO, $G$-AO orders and OL state show off strikingly (see Fig.~\ref{fig:1}).
In fact, in the OL spectra only the acoustic branch is present, whereas in the AO orders ($C$-AO and $G$-AO) spectra the pseudo-optical branch is present as well (compare differences between the FO and AO states in the main text).
Moreover, the optical branch spectral intensities at $\rm{M}=(\pi,\pi,0)$ discriminate between the two different AO orders: while they vanish in the $C$-AO, they diverge in the the $G$-AO case.

Therefore, magnetic RIXS cross section allows one to distinguish between various orbitally ordered phases which are predicted to be stable in KCuF$_3$~\cite{Oles2005}. 
Furthermore, the spectrum of the OL phase (which is not stable in the magnetically ordered phase of KCuF$_3$~\cite{Oles2005}) is strikingly different than the one of the ordered phases which shows that the magnetic RIXS cross section strongly depends on the orbital correlation -- as already discussed in detail in the main part of the paper.

%\bibliography{bib}

\end{document}